\documentclass[cpp,fleqn]{w-art}
\usepackage{times}
\usepackage{w-thm}
\usepackage{cppfig}
\usepackage{amsmath}
\usepackage{amssymb}
\usepackage[]{graphicx}
\begin{document}
\DOIsuffix{theDOIsuffix}
\Volume{42}
\Issue{1}
\Month{01}
\Year{2003}
\pagespan{1}{}
\Receiveddate{}
\Reviseddate{}
\Accepteddate{}
\Dateposted{}



\title[Particle charging]{Towards a microscopic theory of particle charging}


\author[Bronold]{F. X. Bronold\footnote{Corresponding
     author: e-mail:{\sf bronold@physik.uni-greifswald.de}}
     \inst{1}}\address[\inst{1}]{Institut f\"ur Physik,
     Ernst-Moritz-Arndt-Universit\"at Greifswald, D-17489 Greifswald, Germany}
\author[Fehske]{H. Fehske\inst{1}}
\author[Kersten]{H. Kersten\inst{2}}
\address[\inst{2}]{Institut f\"ur Experimentelle und Angewandte Physik, 
     Christian-Albrechts-Universit\"at zu Kiel, D-24098 Kiel, Germany}

\author[Deutsch]{H. Deutsch\inst{1}}
\begin{abstract}
We recently questioned the treatment of a dust particle as a perfect absorber 
for electrons and ions and proposed a surface model for the charge 
of a dust particle in a quiescent plasma which combines the microscopic physics at 
the grain boundary (sticking into and desorption from external surface states) with the 
macrophysics of the discharge (plasma collection fluxes). Within this model the charge 
and partial screening of the particle can be calculated without relying on the 
condition that the total electron collection flux balances on the grain surface the 
total ion collection flux. Grain charges obtained from our approach compared favorably 
with experimental data. The purpose of this paper is to describe our model in more detail,
in particular, the hypotheses on which it is built, contrast it with the standard 
charging models based on flux balancing on the grain surface, and to analyze additional 
experimental data. 
\end{abstract}
\maketitle                   





\section{Introduction}

Dust particles immersed in an ionized gas acquire charges whose signs and 
magnitudes strongly affect the properties of the host plasma as well as 
the properties of the ensemble of dust particles itself. The various 
crystalline and liquid phases of dust particles found in specifically 
designed laboratory experiments are perhaps the most impressive manifestations 
of this effect~\cite{FIK05,Ishihara07}. That dust particles and
host plasma are strongly coupled has been however known for a long time.
Not only from astrophysical plasma environments~\cite{Horanyi96,Mann08} 
but also from processing discharges, where the grains are of course not
the intentional constituents studied for their own sake but contaminants
which need to be controlled because they are detrimental to the overall 
performance of the discharge~\cite{Bouchoule99}.

The charge of a dust particle is an important parameter for the 
quantitative description of dusty plasmas. It has thus been measured in 
a number of experiments~\cite{MTP94,TMP95,WHR95,TLA00,TAA00,SV03,KRZ05}. Considering
the dust particle as a floating electric probe, the interpretation of the
experiments is based on two main assumptions: (i) The grain is a perfect 
absorber for electrons and ions, that is, every electron and ion hitting
the surface of the grain is absorbed and (ii) the quasi-stationary charge 
of the dust particle is the one which balances on the grain surface the 
total electron collection flux with the total ion collection 
flux~\cite{BR59,LP73,Whipple81,Allen92,DPK92,AAA00,KA03}. The charge enters
here through the floating potential of the dust particle which, for a 
spherical particle, can be immediately translated into a charge. In most cases, 
however, the charges obtained from this approach are not in agreement with 
experimental data. Usually, the approximations for the fluxes are blamed for the 
disagreement~\cite{Allen92,AAA00,KA03,LGG01,LGS03,SLR04,KMK06}. We suspect, 
however, that the flux balance condition, as it is currently used, is incomplete 
because it ignores important processes on the grain surface which also affect 
the grain charge. 

First indications that this could be indeed the case came from a phenomenological 
analysis of experimental data based on rate equations for the electron and ion
surface densities of a dust particle~\cite{KDK04}. To provide a more realistic
model for the grain surface, the rate equations contained electron
and ion sticking probabilities, electron and ion desorption 
times, and a electron-ion recombination probability. 
With an appropriate choice of these parameters 
as well as the grain temperature the charges obtained from the quasi-stationarity
of the surface densities were in better agreement with experimental data 
then the ones deduced from the orbital motion limited flux balance condition. 

In a semi-microscopic approach we subsequently pointed out that the charging 
of a dust particle can be interpreted as a physisorption process in the potential 
which builds up in the disturbed region around the grain~\cite{BFK08}. The potential 
is the sum of a short-range polarization potential which enforces the electric boundary 
conditions on the grain surface and, as soon as the grain collected some charge, 
a long-range Coulomb potential. Realizing that the analogy to physisorption suggests, 
on a microscopic scale, a spatial separation of bound electrons and ions, we 
calculated the charge of the grain and its partial screening by balancing, 
on two different effective surfaces, the electron collection flux with the 
electron desorption flux and the ion collection flux with the ion desorption 
flux. The charge of the dust particle thereby obtained is then given by the
number of electrons quasi-bound in the polarization-induced short-range part of 
the particle potential whereas its partial screening is given by the ions quasi-trapped 
in the long-range Coulomb part of the particle potential. 

In this paper we discuss the surface model proposed in~\cite{BFK08}, in particular, 
the hypotheses on which it is built, in more detail and contrast the model with the
standard approaches of calculating the particle charge from flux balancing on the 
grain surface. Thereby we identify certain issues which need to be resolved before 
a more refined microscopic theory of particle charging can be developed. Finally, 
to demonstrate that the surface model in its present form works we analyze further 
experimental data and show that the charges obtained from it are much better than 
the ones deduced from the standard approaches. 

\section{Elementary surface processes}

\begin{figure}[t]
\begin{minipage}{0.45\linewidth}
\includegraphics[width=\linewidth]{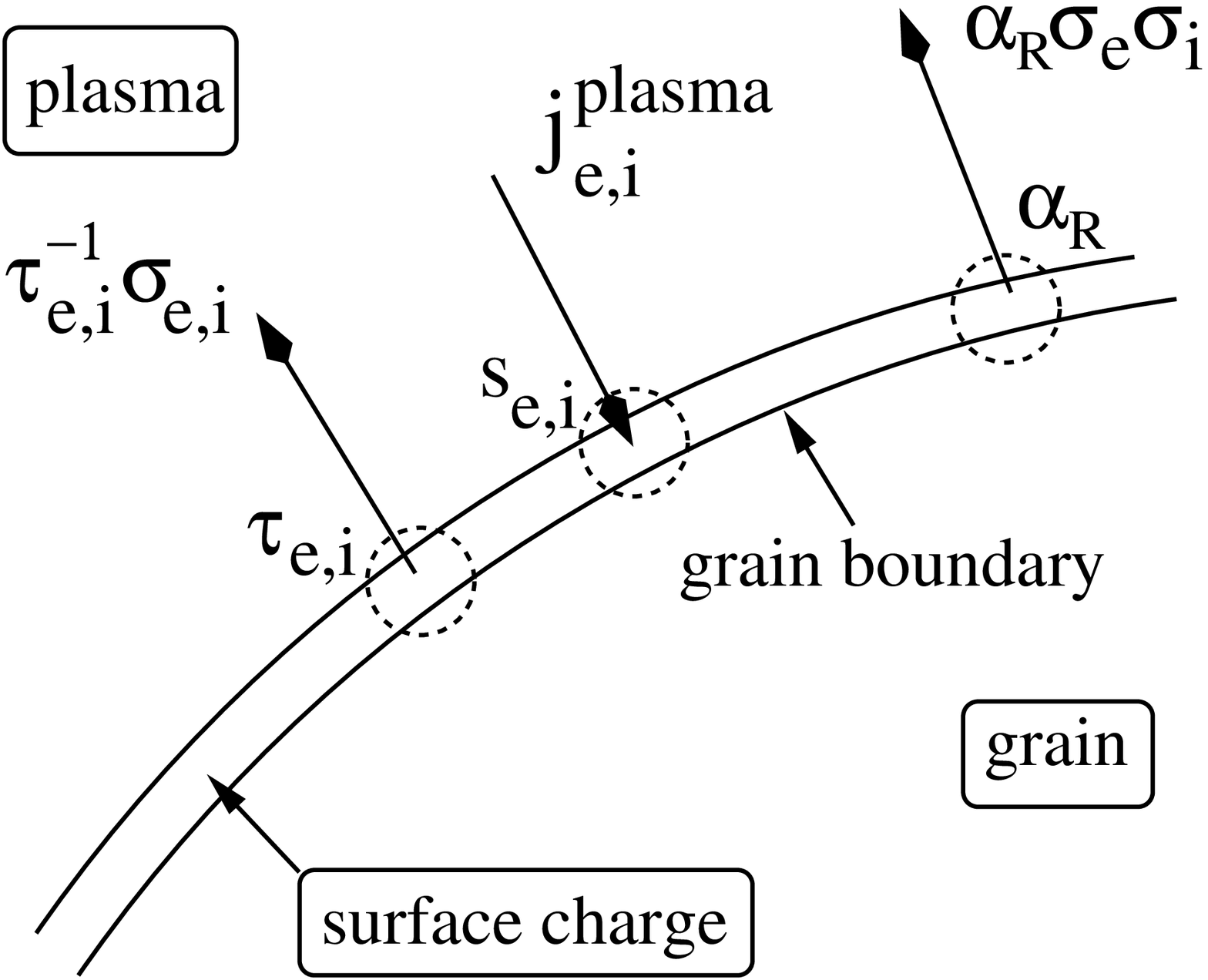}
\end{minipage}\hspace{0.1\linewidth}\begin{minipage}{0.45\linewidth}
\includegraphics[width=\linewidth]{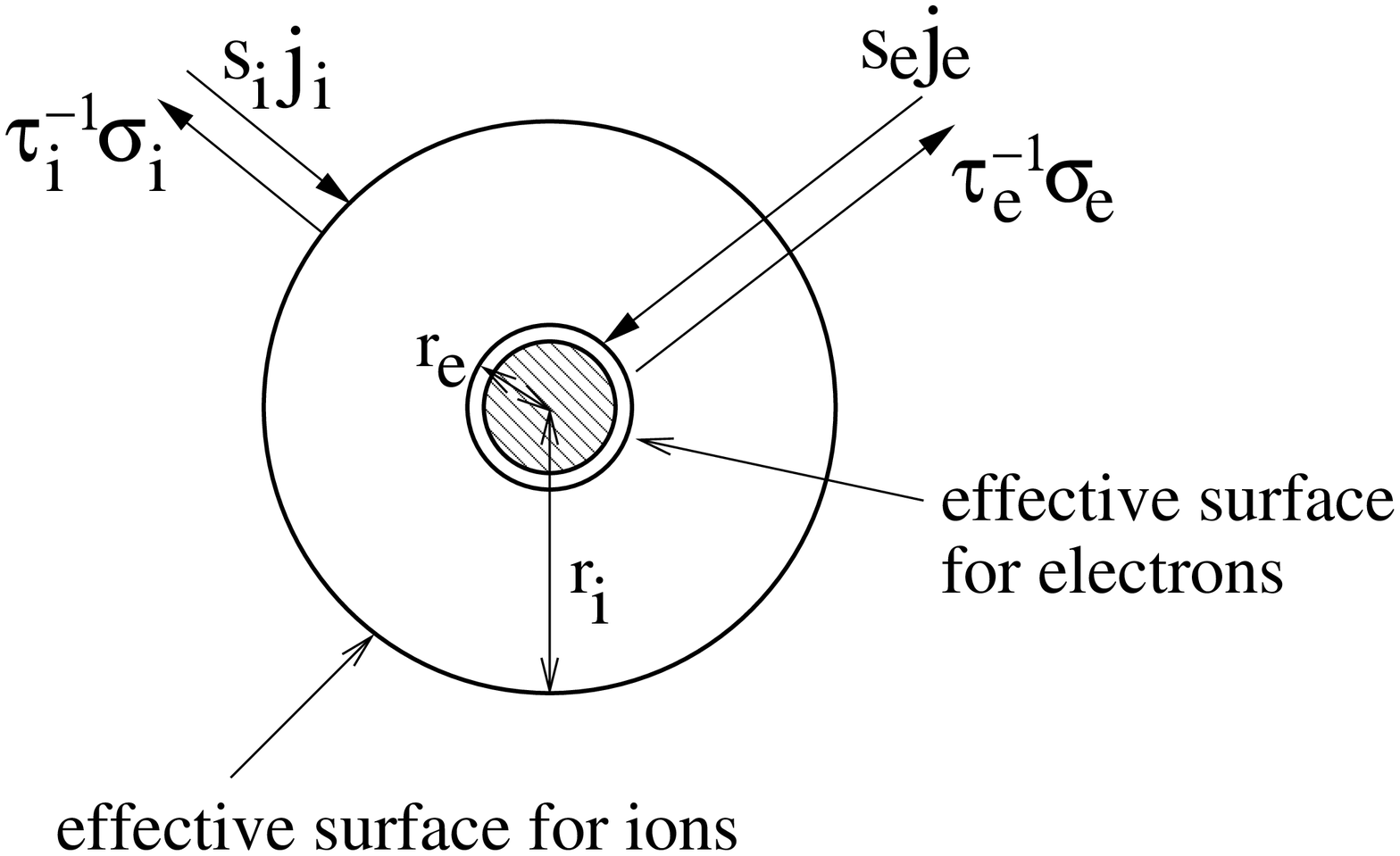}
\end{minipage}
\caption{\label{Cartoons}
Left panel: Generic illustration of elementary surface processes leading
to the formation of surface charges at an inert plasma boundary where 
surface modifications other than charge accumulation do not occur. Right
panel: Surface model proposed in~\cite{BFK08} for the region around a dust 
particle with radius $R$. At quasi-stationarity, the surface charges 
$\sigma_{e,i}$ bound at $r_e\simeq R$ and $r_i \gtrsim r_e$,
respectively, balance the collection fluxes $s_{e,i}j^{\rm plasma}_{e,i}$
with the respective desorption fluxes $\tau_{e,i}^{-1}\sigma_{e,i}$, 
where $s_{e,i}$ and $\tau_{e,i}$ denote, respectively, sticking
coefficients and desorption times.
}
\end{figure}

The microphysics at the grain surface affecting the charge of the grain is schematically
shown in the left panel of Fig~\ref{Cartoons}. Electrons and ions are collected from 
the plasma with collection fluxes $j_{e,i}^{\rm coll}=s_{e,i}j_{e,i}^{\rm plasma}$,
where $s_{e,i}$ are the sticking coefficients and $j_{e,i}^{\rm plasma}$ are the fluxes of
plasma electrons and ions hitting the grain surface. Electrons and ions may thermally desorb
from the grain surface with rates $\tau_{e,i}^{-1}$, where $\tau_{e,i}$ are the desorption 
times. They may also move along the surface with mobilities $\mu_{e,i}$, which in turn may 
affect the probability $\alpha_R$ with which ions recombine with electrons on the surface. 
All these processes occur in a thin layer whose thickness is at most a few microns, that
is, on a scale where the standard kinetic description of the gas discharge based on the 
Boltzmann-Poisson system to which the flux balance condition belongs breaks down.

To put the surface model proposed in~\cite{BFK08} into context and to identify
the assumptions with respect to the surface properties which are usually made 
in the standard calculations of surface charges we cast the elementary surface
processes into rate equations. 

Specifically, we consider a spherical dust particle with radius $R$. The
quasi-stationary charge of the grain is given by (we measure charge in units of $-e$)
\begin{eqnarray}
Z_p=4\pi R^2\sigma_p=4\pi R^2\big[\sigma_{e}-\sigma_i\big]~,
\end{eqnarray}
with electron and ion surface densities, $\sigma_{e,i}$, satisfying the
quasi-stationary ($d\sigma_{e,i}/dt=0$) rate equations~\cite{KDK04},
\begin{eqnarray}
0\!\!\!&=&\!\!\!s_e j^{\rm plasma}_e-\tau_e^{-1}\sigma_e-\alpha_R\sigma_e\sigma_i~,
\label{REqe}\\
0\!\!\!&=&\!\!\!s_i j^{\rm plasma}_i-\tau_i^{-1}\sigma_i-\alpha_R\sigma_e\sigma_i~.
\label{REqi}
\end{eqnarray}
where $j^{\rm plasma}_{e,i}$, $s_{e,i}$, $\tau_{e,i}$, and $\alpha_R$ have been 
introduced above. For simplicity we neglect diffusion along the surface. 

In order to derive the standard criterion invoked to determine the quasi-stationary 
grain charge, we now assume, in contrast to what we do in our model~\cite{BFK08} 
(see below), that both electrons and ions reach the surface of the grain. In that 
case, both Eq.~(\ref{REqe}) and Eq.~(\ref{REqi}) are flux balances on the grain
surface. At quasi-stationarity, the grain is charged to the floating potential. 
In energy units $U=Z_p e^2/R$. Because the grain temperature $kT_p\ll U$ the ion
desorption rate $\tau_i^{-1}\simeq 0$. Equation~(\ref{REqi}) reduces therefore 
to $\alpha_R \sigma_e \sigma_i=s_i j^{\rm plasma}_i$ which transforms Eq.~(\ref{REqe})
into $s_e j^{\rm plasma}_e=s_ij^{\rm plasma}_i+\tau_e^{-1}\sigma_p$ provided
$\sigma_p\simeq\sigma_e$ which is usually the case. In the standard approach the 
grain surface is assumed to be a perfect absorber for electrons and ions. Thus, 
$s_i=s_e=1$ and $\tau^{-1}_e=0$. The quasi-stationary charge $Z_p$ of the grain is 
then obtained from the condition
\begin{eqnarray}
j^{\rm plasma}_e(Z_p)=j^{\rm plasma}_i(Z_p)~,
\label{OrdinaryApp}
\end{eqnarray}
where we explicitly indicated the dependence of the plasma fluxes on the grain charge.

Calculations of the grain charge differ primarily in the approximations made for the plasma
fluxes $j_{e,i}^{\rm plasma}$. For the repelled species, usually collisionless electrons, the flux
can be obtained from Poisson's equation and the collisionless Boltzmann equation, using trajectory
tracing techniques based on Liouville's theorem and momentum and energy conservation~\cite{BR59,LP73,DPK92}.
The flux for the attracted species, usually collisional ions, is much harder to obtain. Unlike the
electron flux, the ion flux depends not only on the field of the macroscopic body but also on
scattering processes due to the surrounding plasma, which throughout we assume to be quiescent.
For weak ion collisionalities the charge-exchange enhanced ion flux model proposed by Lampe and
coworkers~\cite{LGG01,LGS03,SLR04} is often used. Its validity has been however questioned by
Tskhakaya and coworkers~\cite{TTS01,TSS01}. We come back to Lampe and coworkers approach below 
when we discuss representative results of our model.

Irrespective of the approximations made for the plasma fluxes, the standard approaches of calculating
surface charges are based on three assumptions about the surface physics:

(i) Ions and electrons reach the surface, even on the microscopic scale,

(ii) $s_e=s_i=1$ or at least $s_e=s_i$, and

(iii) $\tau_e^{-1}=0$ or at least $\tau_e^{-1}\sigma_e\ll s_ij_i^{\rm plasma}=\alpha_R\sigma_e\sigma_i.$

Hence, electrons and ions hitting the surface of the grain are assumed to be completely absorbed. 
For low-temperature gas discharges, where average electron energies $kT_e\simeq 10~eV$, this 
cannot be the case, however. Permanent implantation of electrons with this energy is very unlikely.
Instead, electrons impinging with a few $eV$ on a surface are either reflected, inelastically
scattered or, when the surface potential supports bound states, temporarily trapped in external 
surface states, with residence times depending on the inelastic coupling of the electrons
to the elementary excitations of the surface and the bulk material. As far as ions approaching 
the dust particle are concerned they can gain at most the floating energy of a few $eV$. They 
are thus also not able to enter the grain and to get permanently stuck.~\footnote{We do not 
account for ion neutralization due to electron capture within the disturbed zone of the grain.}

Based on these considerations, the three assumptions concerning the grain surface made in the 
standard calculations of the grain charge seem to be rather unrealistic. We challenge 
therefore all three of them.

First, electrons and ions should be bound in external surface states. Because of differences 
in the potential energy, mass, and size we expect the spatial extension of the electron 
and ion bound states, and thus the distance of electrons and ions from the grain surface, to
be different. On the microscopic scale, electrons and ions should be spatially separated.

Second, $s_e=s_i$ is quite unlikely. Usually, the sticking coefficient of heavy particles 
is determined by the coupling to vibrational excitations of the surface and the bulk 
material~\cite{KG86,BR92}. This coupling is very strong. Thus, if ions reach the surface, 
as it is conventionally assumed, they would efficiently dissipate energy. The sticking 
probability would be thus large, that is, close to unity. Light particles, 
like electrons, on the other hand, couple only weakly to surface and bulk vibrations. We 
would thus expect $s_e\ll s_i$. In principle the coupling to other elementary excitations
of the grain (plasmons, electron-hole pairs, ...) can compensate for the lack of coupling 
to lattice vibrations but how efficient this coupling really is is not at all obvious. In 
a recent exploratory calculation we found, for instance, that the electron sticking 
coefficient at metallic boundaries arising from the coupling to internal electron-hole 
pairs is also extremely small, at least when the Coulomb interaction between an internal 
and an external electron is realistically screened~\cite{BFD09}.

Third, if ions and electrons are indeed spatially separated, the two rate equations 
(without the recombination term) should be in fact interpreted as flux balances on two 
different effective surfaces as shown in the right panel of Fig.~\ref{Cartoons}. In that 
case, $\alpha_R\sigma_i\sigma_e\ll \sigma_{e,i}/\tau_{e,i}$ and the surface charge $Z_p$ 
would be determined by balancing on the grain surface the electron desorption flux,
$\tau_e^{-1}\sigma_e$, with the electron collection flux, $s_ej_e^{\rm plasma}$. The
corresponding balance of ion fluxes, to be taken on an effective surface surrounding
the grain, would then yield a partial screening charge $Z_i$. Within this scenario, we
would thus obtain
\begin{eqnarray}
Z_p&=&4\pi r_e^2\cdot (s\tau)_e\cdot j_e^{\rm plasma}(Z_p)~,
\label{OurZp}\\
Z_i&=&4\pi r_i^2\cdot (s\tau)_i\cdot j_i^{\rm plasma}~,
\label{OurZi}
\label{Charge}
\end{eqnarray}
with $r_e\simeq R$ and $r_i\gtrsim r_e$, which are the equations proposed 
in~\cite{BFK08}. The sticking into and desorption from external surface states is 
here encoded in the products $(s\tau)_{e,i}$. For electrons the product depends only 
on material parameters but for ions it also depends on plasma parameters.

\section{Surface states}

The starting point of our surface model was a quantum mechanical investigation of the
bound states of a negatively charged particle in a gas discharge. More specifically, 
we considered the static interaction between an electron (ion) with charge $-e$ ($+e$) 
and a spherical particle with radius $R$, dielectric constant $\epsilon$, and charge 
$Q_p=-eZ_p$. 

As mentioned before, the interaction potential contains a polarization-induced part, 
arising from the electric boundary conditions at the grain surface, and a Coulomb 
tail due to the particle's charge~\cite{DS87,Boettcher52}. For both terms we adopted
the simplest approximations. The Coulomb part is then the potential of a point 
charge whose magnitude is the charge of the grain and the polarization part is the 
classical image potential. More sophisticated treatments taking, for instance, 
the finite velocity of the approaching electron (ion) and the nonlocality of the 
polarization potential at short distances into account are possible but not 
needed at this stage of the investigation.

\begin{figure}[t]
\begin{center}
\includegraphics[width=0.7\linewidth]{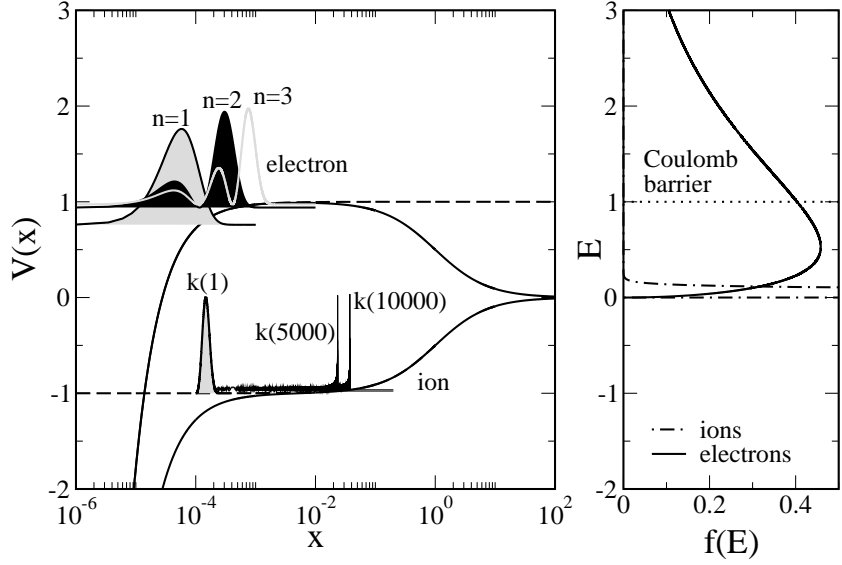}
\caption{\label{Potential}
Left panel: Potential energy for an electron (ion) in the
field of a MF particle ($R=4.7~\mu m$, $Z=6800$)~\cite{KDK04,Melzer97}
and representative probability distributions, $|u(x)|^2$,
shifted to the binding energy and maxima normalized to one.
Dashed lines denote the potentials used in the
Schr\"odinger equations. Note, the finite ion radius $\rho_i^{size} \simeq \AA$
forces the ion wavefunctions to vanish at $x\le x_i^{size}\simeq 10^{-4}$.
Right panel: Bulk energy distribution functions for the
discharge hosting the particle~\cite{KDK04,Melzer97}.
}
\end{center}
\end{figure}

Measuring distances from the grain surface in units of $R$ and energies in units
of $U$, the interaction energy at $x=r/R-1>x_b$, where $x_b$ is a lower
cut-off, below which the grain boundary cannot be described as a perfect surface
anymore, reads
\begin{eqnarray}
V_{e,i}(x)&=&\pm\frac{1}{1+x}-\frac{\xi}{x(1+x)^2(2+x)}
\nonumber\\
&\simeq&\left\{\begin{array}{ll}
1-\xi/2x & \mbox{electron}\\
-1/(1+x) & \mbox{ion}
\end{array}\right.
\label{V(x)}
\end{eqnarray}
with $\xi=(\epsilon-1)/2(\epsilon+1)Z_p$.

The second line in Eq.~(\ref{V(x)}) is an approximation which describes
the relevant parts of the potential very well and yet permits an analytical
calculation of the surface states. In Fig.~\ref{Potential} we plot
$V_{e,i}(x)$ for a melamine-formaldehyde (MF) particle ($\epsilon=8$,
$R=4.7~\mu m$, and $Z_p=6800$) embedded in a helium discharge with
plasma density $n_e=n_i=0.62\times 10^9~cm^{-3}$, ion temperature
$k_BT_i=0.04~eV$, and electron temperature $k_BT_e=2.2~eV$~\cite{KDK04,Melzer97}.
From the electron energy distribution, $f_{e}(E)$, we see that
the discharge contains enough electrons which can overcome the Coulomb
barrier of the particle which is the floating energy $U$. These electrons 
may get bound in the polarization-induced short-range part of the potential, 
well described by the approximate expression, provided they can get rid of 
their kinetic energy. Ions, on the other hand, being cold
(see $f_i(E)$ in Fig.~\ref{Potential}) and having a finite
radius $\rho^{size}_i/R=x_i^{size}\simeq 10^{-4}$, cannot explore the
polarization potential at short distances~\footnote{We treat the ion as a 
structureless rigid sphere.}. For them, the long-range Coulomb tail is most 
relevant, which is again well described by the approximate expression.

Writing for the electron eigenvalue $\varepsilon^{e}=1-\alpha_e\xi/4k^2$
with $\alpha_e=(\epsilon-1)R/4(\epsilon+1)a_B$ and for the ion eigenvalue
$\varepsilon^{i}=-\alpha_i/2k^2$ with $\alpha_i=m_iRZ_p/m_ea_B$, where
$a_B$ is the Bohr radius and $m_e$ and $m_i$ are the electron and ion mass,
respectively, the radial Schr\"odinger equations for the approximate potentials
read
\begin{eqnarray}
\frac{d^2u^{e,i}}{dx^2}+\bigg[-\frac{\alpha_{e,i}^2}{k^2}+\tilde{V}_{e,i}(x)
-\frac{l(l+1)}{(1+x)^2}
\bigg]u^{e,i}=0
\label{SE}
\end{eqnarray}
with $\tilde{V}_e(x)=2\alpha_e/x$ and $\tilde{V}_i(x)=2\alpha_i/(1+x)$.
For bound states, the wavefunctions have to vanish for $x\rightarrow\infty$.
The boundary condition at $x_b$ depends on the potential for
$x\le x_b$, that is, on the surface barrier (which is
different for electrons and ions). Matching the solutions for $x<x_b$
and $x>x_b$ at $x=x_b$ leads to a secular equation for $k$. For our
purpose, it is sufficient to take the simplest model for the
barrier: $\tilde{V}_{e,i}(x\le x_b)=\infty$ with $x_b=0$ for electrons
and $x_b=x_i^{size}$ for ions.

The electron Schr\"odinger equation is then equivalent to the Schr\"odinger
equation for the hydrogen atom and $k$ is an integer $n$. Because (for
bound electrons) $x\ll 1$ and $\alpha_e\gg 1$, the centrifugal
term is negligible. Hence, we consider only states with
$l=0$. The eigenvalues are then
$\varepsilon^e_n=1-\alpha_e\xi/4n^2$
and the wavefunctions read
\begin{eqnarray}
u_{n,0}^e(x)\sim v_{n,0}(\bar{z})
=\bar{z}\exp(-\bar{z}/2)(-)^{n-1}(n-1)!L^{(1)}_{n-1}(\bar{z})
\label{vn0}
\end{eqnarray}
with $\bar{z}=2\alpha_ex/n$ and $L^{(1)}_n(\bar{z})$ associated Laguerre
polynomials.

The probability densities $|u_{n,0}^e(x)|^2$ for the first three electron
surface states are plotted in Fig.~\ref{Potential}. As can be seen, they 
are very close to the surface. In physical units, the electron surface 
states are at most a few $\AA$ngstroms away from the grain boundary. At
these distances, the spatial variation of $V_e(x)$ is comparable to the
de-Broglie wavelength of electrons approaching the particle.
More specifically, for $k_BT_e=2.2~eV$, $\lambda_e^{dB}/R\simeq |V_e/V_e'|
\simeq 10^{-4}$. Hence, the trapping of electrons at the surface of the 
particle is a quantum-mechanical effect beyond the Boltzmann-Poisson 
description of the plasma-grain interaction. 

The solutions of the ion Schr\"odinger equation are Whittaker functions,
$u^i_{k,l}(x)=W_{k,l+1/2}(\bar{x})$ with $\bar{x}=2\alpha_i(1+x)/k$ and
$k$ determined from $u^i_{k,l}(x_i^{size})=0$. However, since
$\alpha_i\gg 1$ and $k\gg 1$, it is very hard to work directly with 
$W_{k,l+1/2}(\bar{x})$. It is easier to use the method of comparison 
equations~\cite{Richardson73} and to construct uniform approximations for 
$u^i_{k,l}(x)$ with the radial Schr\"odinger equation for the hydrogen atom 
as a comparison equation. The method can be applied for any $l$. Here
we give only the result for $l=0$:
\begin{eqnarray}
u_{k,0}^i(x)\sim v_{n,0}(\bar{z})/\sqrt{dz/dx}
\end{eqnarray}
with $v_{n,0}(\bar{z})$ defined in Eq.~(\ref{vn0}) and $\bar{z}=2\alpha_i z(x)/n$. 
The mappings $z(x)$ and $k(n)$ can be constructed from the phase-integrals of the
two ion Schr\"odinger equations~\cite{Richardson73}.

In Fig.~\ref{Potential} we show $|u_{k,0}^i(x)|^2$ for $k(1)$, $k(5000)$ and
$k(10000)$. Note, even the $k(10000)$ state is basically at the bottom
of the potential. This is a consequence of $\alpha_i\gg 1$ which leads
to a continuum of bound states below the ion ionization threshold at $\varepsilon=0$.
We also note that $|u_{k(n),0}^i(x)|^2$ peaks for $n\gg 1$ just below
the turning point. Hence, except for the lowest ion surface states, which we 
expect to be of little importance, ions are essentially trapped in classical
orbits deep in the disturbed region of the grain. This will be also the
case for $l>0$. That ions behave classically is not unexpected because
for $k_BT_i=0.04~eV$ their de-Broglie wavelength is much smaller then
the scale on which the potential varies for $x>10^{-3}$:
$\lambda_i^{dB}/R\simeq 10^{-5}\ll |V_i/V_i'|\simeq 1$. Thus, the
interaction between ions and the dust particle is classical and can be 
analyzed with Boltzmann-Poisson equations.

Nevertheless it is also possible to describe this interaction quantum-mechanically. We
anticipate even a quantum-mechanical approach, based on the method of comparison
equations, which is an asymptotic technique well suited for the semiclassical 
domain we are interested in, to be rather useful in this respect. 
In fact, many years ago a wave-mechanical description of the collisionless ion 
dynamics around electric probes has been pursued by Liu~\cite{Liu69} but he found
no followers.

\section{Charging model}

Using the results of the previous section, a model for the grain charge taking
surface states into account can be constructed as follows. Within the disturbed region 
of the particle, the density of free electrons (ions) is much smaller than the density 
of bound electrons (ions). In that region, the quasi-stationary charge (again in 
units of $-e$) is thus approximately given by
\begin{eqnarray}
Z(x)=4\pi R^3\!\int_{x_b}^x\!\!dx' \big(1+x'\big)^2 \bigg[n^b_e(x')-n^b_i(x')
\bigg]
\label{Zintegral}
\end{eqnarray}
with $x<\lambda^D_i=\sqrt{kT_i/4\pi e^2 n_i}$, the ion Debye length, which we take as
an upper cut-off, and $n^b_{e,i}$ the density of bound electrons and ions. For the
plasma parameters used in Fig.~\ref{Potential}, $\lambda^D_i\simeq 60\mu m$.
The results for the surface states presented above suggest to express the density
of bound electrons by an electron surface density:
\begin{eqnarray}
n^b_e(x)\simeq\sigma_e\delta(x-x_e)/R
\label{nb}
\end{eqnarray}
with $x_e\simeq x_b\simeq 0$ and $\sigma_e$ the quasi-stationary solution of
Eq.~(\ref{REqe}) without the recombination term. Equation~(\ref{REqe}) is thus 
still a rate equation on the grain surface. We will argue below that once the
grain has collected some negative charge, not necessarily the quasi-stationary 
one, there is a critical ion orbit at $x_i\sim 1-10 \gg x_e$ which prevents ions 
from hitting the particle surface. Thus, the particle charge obtained from 
Eq.~(\ref{Zintegral}) is simply $Z_p\equiv Z(x_e<x<x_i)$. Inserting Eq. (\ref{nb}) 
into Eq.~(\ref{Zintegral}) and integrating up to $x$ with $x_e<x<x_i$ leads to 
Eq.~(\ref{OurZp}), the expression for the particle charge deduced from the rate 
equations (\ref{REqe}) and (\ref{REqi}) under the assumption that ions do not reach 
the grain surface on the microscopic scale.

\begin{figure}[t]
\begin{center}
\begin{minipage}{0.4\linewidth}
\includegraphics[width=\linewidth]{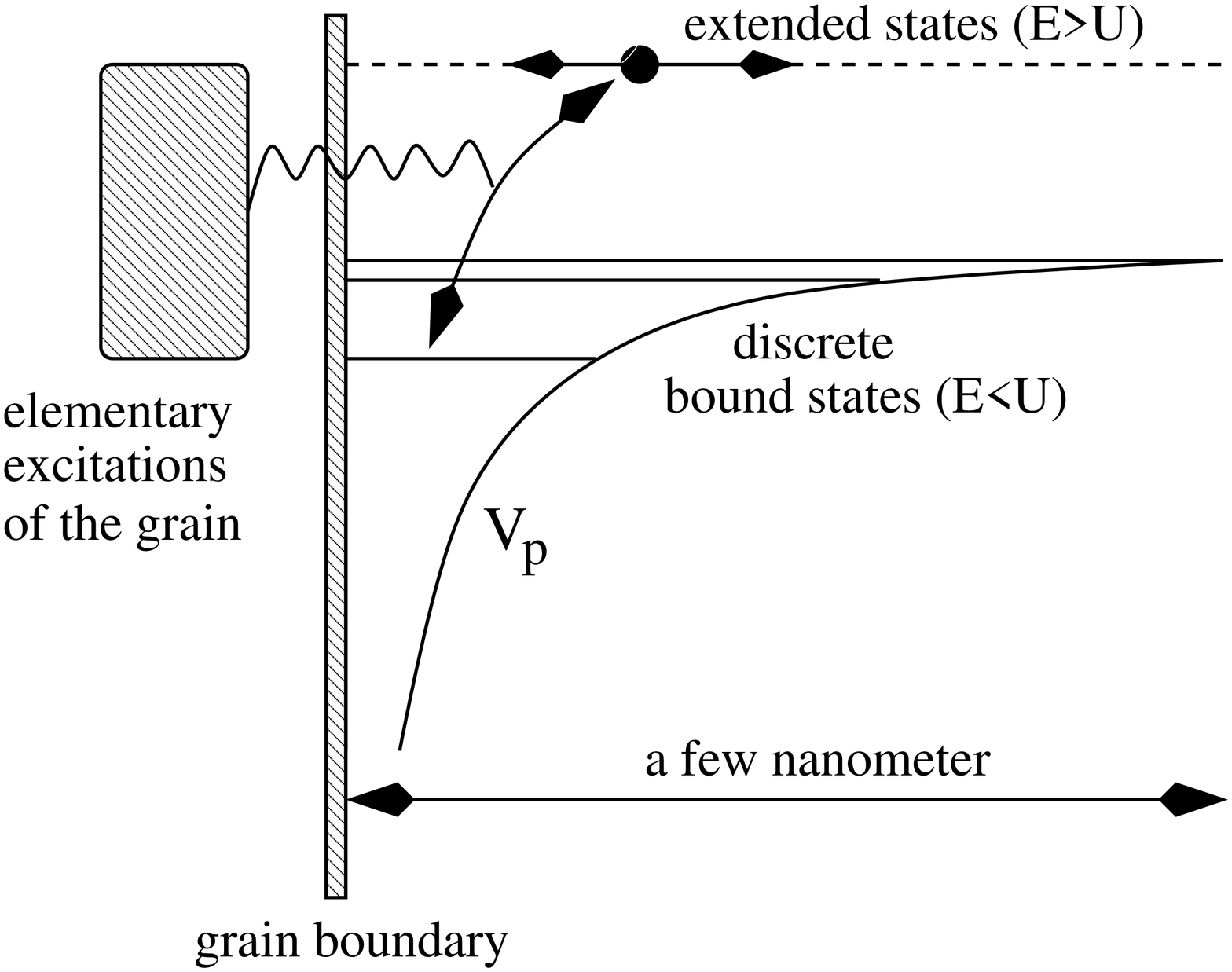}
\end{minipage}\hspace{0.05\linewidth}\begin{minipage}{0.4\linewidth}
\includegraphics[width=\linewidth]{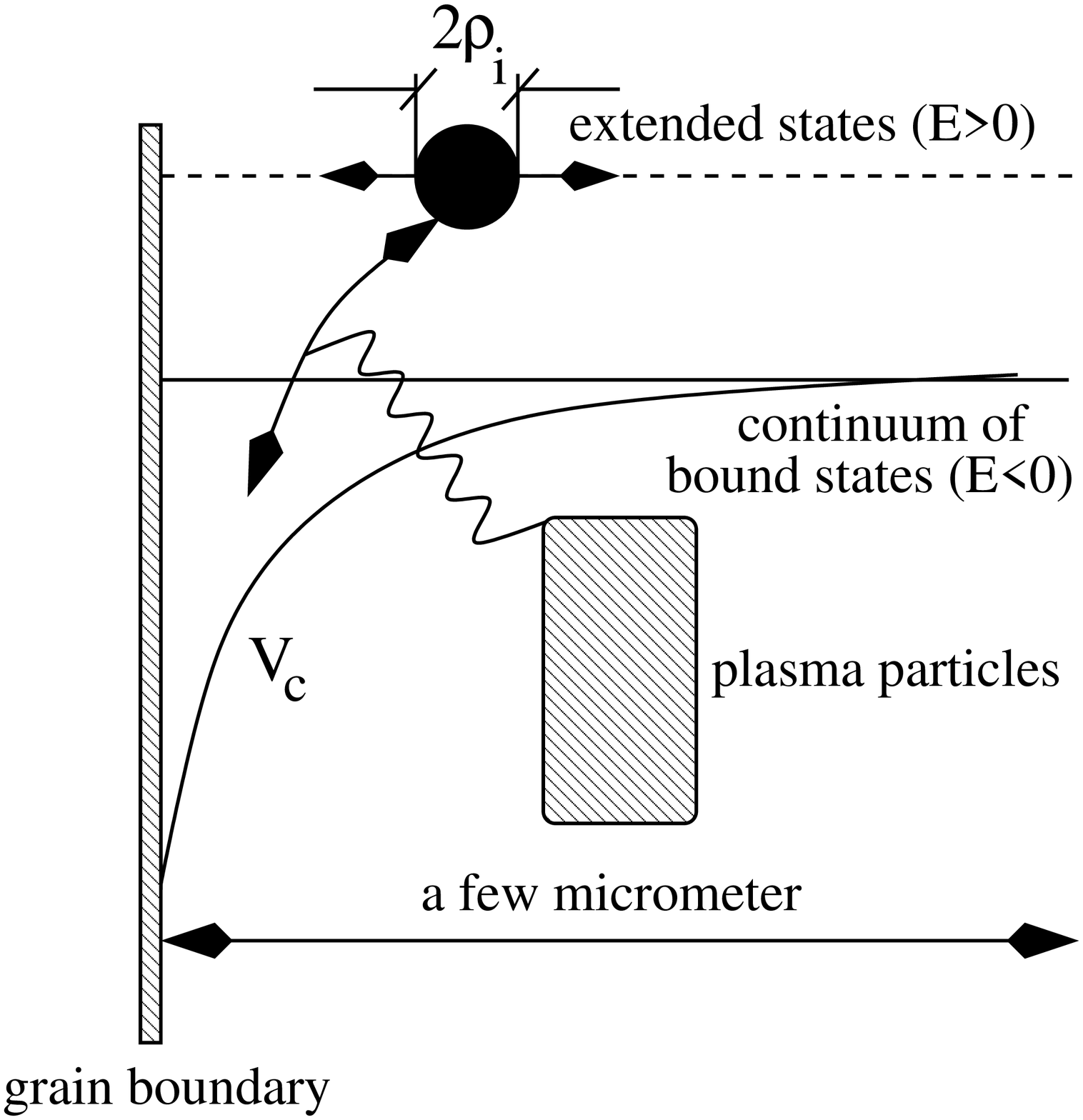}
\end{minipage}
\caption{\label{Erelax}
Left panel: Schematic illustration of electron energy relaxation at the 
grain boundary. The electron looses (gains) energy due to creation (annihilation) 
of elementary excitations of the grain. Due to these processes it may get trapped 
in (escape from) the discrete bound states of the short-range polarization potential.
Right panel: Schematic illustration of ion energy relaxation in the vicinity 
of the grain surface. The ion looses (gains) energy due to collisions with other
plasma particles (electrons, ions, neutrals). It is due to these processes that 
an ion may get trapped in (escape from) the continuum of bound states of the 
long-range Coulomb potential.
}
\end{center}
\end{figure}

For an electron to get stuck at (to desorb from) a surface it has to loose (gain)
energy at (from) the surface~\cite{KG86}. This can only occur through inelastic
scattering with the elementary excitations of the grain schematically shown in 
the left panel of Fig.~\ref{Erelax}. To calculate the product $(s\tau)_e$ requires
therefore a microscopic model for the electron-grain interaction. First steps
in this direction were taken in~\cite{BFD09}. 

In Ref.~\cite{BFK08} we invoked however the phenomenology of reaction rate theory 
to approximate $(s\tau)_e$. Specifically, we assumed that electrons with rather 
low and rather high energies are, respectively, reflected by the Coulomb and 
surface barrier of the particle and that sticking (desorption) primarily affects
electrons with energies slightly above $U$. After overcoming the Coulomb barrier 
this group of electrons is almost in equilibrium with the surface electrons. We 
can thus apply the Lennard-Jones-Devonshire formula~\cite{KG86,LJD36} to obtain, 
\begin{eqnarray}
(s\tau)_e=
\frac{h}{k_BT_p}\exp\bigg[\frac{E_e^d}{k_BT_p}\bigg]~,
\label{stau}
\end{eqnarray}
where $h$ is Planck's constant, $T_p$ is the grain temperature, and $E_e^d$ is the
electron desorption energy, that is, the binding energy of the surface state from
which desorption most likely occurs. The great virtue of this equation is that
it relates a combination of kinetic coefficients, which depend on the details
of the inelastic (dynamic) interaction, to an energy, which can be deduced from
the static interaction alone. Kinetic considerations are thus reduced to a minimum.
They are only required to identify the relevant temperature and the state from which
desorption most probably occurs. 

Equation~(\ref{OurZp}) is a self-consistency equation for $Z_p$. Combined with
Eq.~(\ref{stau}), and approximating the electron plasma flux $j_e^{\rm plasma}$ 
by the orbital motion limited flux,
\begin{eqnarray}
j_e^{\rm OML}=n_e\sqrt{\frac{k_BT_e}{2\pi m_e}}\exp\bigg[\frac{-Z_pe^2}{Rk_BT_e}\bigg]~,
\label{jeOML}
\end{eqnarray}
which is reasonable, because, on the plasma scale, electrons are repelled from the
grain surface, the grain charge is given by
\begin{eqnarray}
Z_p=4\pi R^2\frac{h}{k_BT_p}\exp\bigg[\frac{E_e^d}{k_BT_p}\bigg]j_e^{\rm OML}(Z_p)~.
\label{Zpfinal}
\end{eqnarray}
Thus, in addition to the plasma parameters $n_e$ and $T_e$, the charge depends on
the surface parameters $T_p$ and $E_e^d$.

Without a microscopic theory for the inelastic electron-grain interaction, a plausible
estimate for $E_e^d$ has to be found from physical considerations alone. Since by 
necessity the electron comes very close to the grain surface it will strongly couple
to elementary excitations of the grain (see Fig~\ref{Potential} and the left panel 
of Fig.~\ref{Erelax}). Depending on the material these may be bulk or surface phonons, 
bulk or surface plasmons, or internal electron-hole pairs. For any realistic surface 
barrier, where the electron wavefunction leaks into the solid, the electron will thus 
quickly relax to the lowest surface bound state~\footnote{The microscopic model for 
electron energy relaxation at metallic boundaries employed in~\cite{BFD09} works also 
for an infinitely high barrier.}. The $n=1$ state for the infinitely high 
barrier is an approximation to that state. Thus, it is reasonable to expect
\begin{eqnarray}
E_e^d\simeq (1-\varepsilon^e_1)U=\frac{R_0}{16}\bigg(\frac{\epsilon-1}{\epsilon+1}\bigg)^2~,
\end{eqnarray}
where $\varepsilon^e_1$ is the lowest eigenvalue of the electronic Schr\"odinger
equation. For an MF particle with $\epsilon=8$ this leads to $E_e^d\simeq 0.5eV$. The particle 
temperature cannot be determined in a simple way. It depends on the balance of heating 
and cooling fluxes to-and-fro the particle and thus on additional surface parameters~\cite{SKD00}. 
We use $T_p$ therefore as an adjustable parameter. To reproduce, for instance, with 
Eq.~(\ref{Zpfinal}) the charge of the particle in Fig.~\ref{Potential}, $T_p=395~K$ implying 
$(s\tau)_e\simeq 10^{-6}~s$.

Equation~(\ref{Zpfinal}) depends on the assumption that once the particle is negatively
charged ions are trapped far away from the grain surface. Indeed, treating trapping
of ions in the field of the grain as a physisorption process suggests this assumption, 
which is perhaps counter-intuitive. Provided the ion remains intact on its way 
towards the grain surface (no ion neutralization in the disturbed zone of the grain
due to electron capture) it gets bound to the grain only when it looses energy. 
Because of its low energy and the long-range attractive ion-grain interaction, the ion 
will be initially bound very close to the ion ionization threshold (see Fig.~\ref{Potential}). 
The coupling to the elementary excitations of the grain is thus negligible and only collisions
with other plasma particles are able to push the ion from an extended state with positive 
energy to a bound state with negative energy and then from a given bound state to a lower
one (see the illustration in the right panel of Fig.~\ref{Erelax}). Since the interaction is
classical, collisions, for instance, charge-exchange scattering between ions and atoms, 
act like a random force. Ion energy relaxation can thus be envisaged as a destabilization
of orbits. This is in accordance to what Lampe and coworkers assume~\cite{LGG01,LGS03,SLR04}.
In contrast to them, however, we~\cite{BFK08} expect orbits whose spatial extension is
smaller than the scattering length to be stable because the collision probability during 
one revolution becomes vanishingly small. For a circular orbit, a rough estimate for the 
critical radius is $r_i=R(1+x_i)=(2\pi\sigma_{\rm cx} n_g)^{-1}$ which leads to
$x_i\simeq 5\gg x_e \simeq 0$ when we use the parameters of the helium discharge of
Fig.~\ref{Potential} and $\sigma_{\rm cx}=0.32\times 10^{-14}~cm^2$ which is the measured
cross section at $0.3~eV$~\cite{Helm77,SLB79}.

Indeed, Lampe and coworkers approach~\cite{LGG01,LGS03,SLR04} shows a pile-up 
of trapped ions in a shell of a few $\mu m$ radius enclosing the grain. They would however
not expect a relaxation bottleneck. This point can be only clarified with a detailed 
investigation of the ion dynamics and kinetics in the disturbed region of the grain taking 
the complete kinematics of charge-exchange collisions encoded in the differential collision 
cross section and the centrifugal barriers separating bound from unbound ion motion into 
account. In fact, Lampe and coworkers neglect the momentum transfer during a charge-exchange 
collision as well as the barriers. Tskhakaya and coworkers~\cite{TTS01,TSS01} on the 
other hand pointed out that the latter could severely overestimate the collision-enhanced 
ion flux. In reality, this flux, they claim, is much smaller than the one obtained by Lampe
and coworkers. If this is indeed the case, the charges obtained from the collision-enhanced 
ion flux model would be much closer to the orbital-motion limited ones and thus far away from
the experimentally measured charges (see next section).

\begin{table}
\caption{\label{HePara} Plasma parameters~\cite{KDK04,Melzer97}
used to obtain the results plotted in Fig.~\ref{Melzer}.}
\begin{tabular}{lllll}
$P[W]$ & $p[Pa]$ & $n_e,n_i[10^9 cm^{-3}]$ & $kT_e[eV]$ & $kT_i[eV]$ \\\hline
5 & 102 & 0.33 & 1.5 & 0.030\\
12 & 102 & 0.62 & 2.2 & 0.040\\
20 & 102 & 0.9 & 1.7 & 0.035\\
30 & 102 & 1.2 & 1.28 & 0.036\\
40 & 102 & 1.5 & 1.2 & 0.038\\
50 & 102 & 1.4 & 1.0 & 0.039\\
60 & 102 & 1.6 & 0.8 & 0.040\\\hline
12 & 30 & 0.26 & 2.2 & 0.030\\
12 & 40 & 0.4 & 2.2 & 0.030\\
12 & 60 & 0.48 & 2.2 & 0.030\\
12 & 80 & 0.52 & 2.2 & 0.036\\
12 & 102 & 0.62 & 2.2 & 0.040\\
12 & 120 & 0.8 & 2.2 & 0.040\\\hline
\end{tabular}
\end{table}

In~\cite{BFK08} we pushed the assumption of a critical ion orbit to its limit and approximated 
the density $n_i^b$ of ions accumulating in the disturbed region of the grain, and being 
responsible for the partial screening of the grain charge, by a surface density $\sigma_i$ 
which balances at $x_i$ the ion charging flux with the ion desorption flux (see right panel of
Fig.~\ref{Cartoons}). Mathematically, this gives rise to a rate equation similar to~(\ref{REqi}),
but without the recombination term and interpreted as a rate equation at $r=r_i$. At 
quasi-stationarity, the ion surface density is thus $\sigma_i=(s\tau)_i j_i$. Although 
Eq.~(\ref{stau}) assumes elementary excitations of the grain to be responsible for sticking 
and desorption we expect a similar expression (with $E_e^d$, $T_p$ replaced by $E_i^d$, $T_g$) 
to control the density of trapped ions. Equation~(\ref{Zintegral}) leads then to 
$Z(x_i<x<\lambda^D_i)=Z_p-Z_i$ with
\begin{eqnarray}
Z_i=4\pi R^2(1+x_i)^2
\frac{h}{k_BT_g}\exp\bigg[\frac{E_i^d(Z_p)}{k_BT_g}\bigg]j^B_i~
\label{Zion}
\end{eqnarray}
the number of trapped ions where we assumed that the critical ion orbit, which is near the 
sheath-plasma boundary, is fed by the Bohm ion flux $j^B_i=0.6n_i\sqrt{k_BT_e/m_i}$. 

The ion desorption energy is the negative of the binding energy of the critical orbit,
\begin{eqnarray}
E_i^d(Z_p)=-V_i(x_i)U=4\pi\sigma_{\rm cx}a_B n_g Z_pR_0~,
\end{eqnarray}
and depends strongly on $Z_p$ and $x_i$. For the situation shown in Fig.~\ref{Potential} 
we obtained, for instance, using $T_g=T_p=395~K$, the particle temperature which 
reproduces $Z_p=6800$, $E_i^d\simeq 0.37~eV$ and $(s\tau)_i\simeq 0.6\times 10^{-8}~s$.
The ion screening charge is then $Z_{i}\simeq 148\ll Z_p$ which is the order of magnitude
expected from molecular dynamics simulations~\cite{CK94}. Thus, even when the particle
charge is defined by $Z(x_i<x<\lambda^D_i)$ it is basically given by $Z_p$.

\section{Results}

\begin{figure}[t]
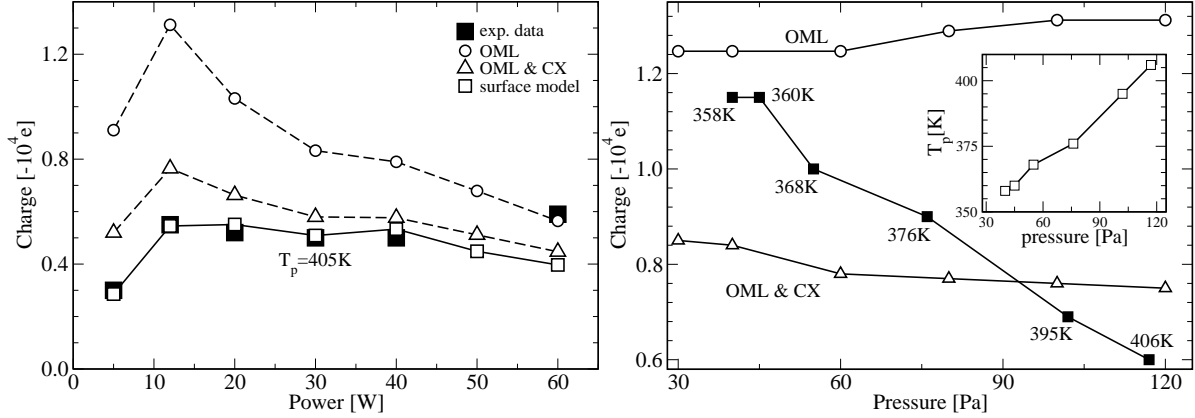

\begin{center}
\begin{minipage}{0.5\linewidth}
\includegraphics[width=\linewidth]{Fig4l.eps}
\end{minipage}\begin{minipage}{0.5\linewidth}
\includegraphics[width=\linewidth]{Fig4r.eps}
\end{minipage}
\caption{\label{Melzer}
Left and right panel show, respectively, the power and pressure dependence of
the charge of a  MF particle ($R=4.7~\mu m$) in the helium discharge of Ref.~\cite{Melzer97}.
The particle temperatures reproducing the experimental data (filled squares) are
indicated and the charges obtained from $j_e^{\rm OML}=j_i^{\rm OML}$ (OML) and
$j_e^{\rm OML}=j_i^{\rm OML}+j_i^{\rm CX}$ (OML~$\&$~CX) with
$\sigma_{\rm cx}=0.3\times 10^{-14} cm^2$~\cite{Helm77,SLB79}, which is the measured
cross section at $0.3~eV$, are also shown. The predicted increase of the particle
temperature with pressure is plotted in the inset of the right panel.
}
\end{center}
\end{figure}

We now use Eq.~(\ref{Zpfinal}) to calculate for various experimental set-ups the
charge of a dust particle. In all cases the plasma parameters are known. The only
free parameter is thus the particle temperature. Although there exist optical 
methods to measure the temperature of dust particles~\cite{MBK08} they have not 
yet been used in conjunction with charge measurements. In view of the importance 
of the particle temperature for the surface model, we hope that in the near 
future an experimental group can be found to perform such measurements. 

In addition to the results obtained from the surface model we also plot charges 
deduced from balancing the orbital motion limited electron flux $j_e^{\rm OML}$ 
given in Eq.~(\ref{jeOML}) with, respectively, the orbital motion limited ion flux, 
\begin{eqnarray}
j_i^{\rm OML}=n_i\sqrt{\frac{k_BT_i}{2\pi m_i}}\bigg[1+\frac{Z_pe^2}{R k_BT_i}\bigg]~,
\end{eqnarray}
and the collision-enhanced ion flux (see~\cite{LGG01,LGS03} and also~\cite{KRZ05}),
\begin{eqnarray}
j_i^{\rm OML}+j_i^{\rm CX}=n_i\sqrt{\frac{k_BT_i}{2\pi m_i}}\bigg[1+\frac{Z_pe^2}{R k_BT_i}
+0.1\frac{\lambda^D_i}{l_{\rm cx}}
\bigg(\frac{Z_pe^2}{R k_BT_i}\bigg)^2\bigg]~,
\end{eqnarray}
where $l_{\rm cx}=(\sigma_{\rm cx} n_g)^{-1}$ is the scattering length,
$\sigma_{\rm cx}$ is the charge-exchange cross section, and $n_g=p/k_BT_g$ is the gas
density.

We start with MF particles confined in a helium discharge. In Fig.~\ref{Melzer} we show 
the power and pressure dependence of the charge of a particle with $R=4.7~\mu m$ 
at rest in the helium discharge of Ref.~\cite{Melzer97}, the plasma parameters of which 
are given in Table~\ref{HePara}. The $p=102~Pa$ data point of the $P=12~W$ run served as 
an illustration in Fig.~\ref{Potential}. Using the parameters of Table~\ref{HePara} we 
calculated from Eq.~(\ref{Zpfinal}) for each data point $T_p$ such that $Z_p(T_p)=Z_{\rm exp}$. 
The power dependence of the charges, shown in the left panel of Fig.~\ref{Melzer}, 
could be reproduced by a single particle temperature $T_p=405~K$ while the pressure 
dependence shown in the right panel required to adjust for each pressure the particle 
temperature. Assuming $T_p$ to scale with the gas temperature $T_g$, the predicted 
particle temperature plotted in the inset of the right panel is in accordance with what 
we would expect from the pressure dependence of $T_g$ in noble gas discharge~\cite{SKD00} 
indicating that our approach gives physically consistent results. For comparison we also 
plot the particle charges deduced, respectively, from $j_e^{\rm OML}=j_i^{\rm OML}$ and 
$j_e^{\rm OML}=j_i^{\rm OML}+j_i^{\rm CX}$. Obviously, the agreement with the data is 
not very good.

\begin{figure}[t]
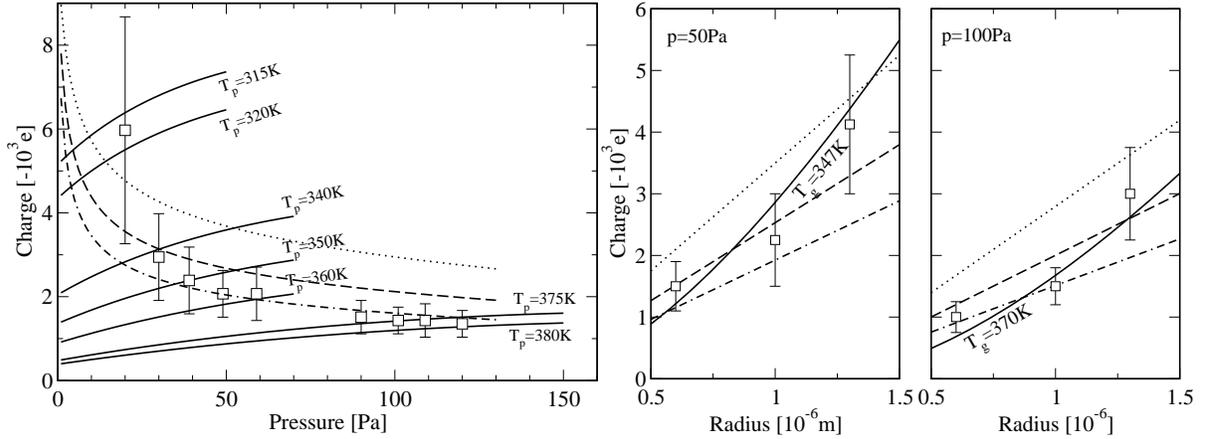

\begin{center}
\begin{minipage}{0.5\linewidth}
\includegraphics[width=\linewidth]{Fig5l.eps}
\end{minipage}\begin{minipage}{0.5\linewidth}
\includegraphics[width=\linewidth]{Fig5r.eps}
\end{minipage}
\caption{\label{Khrapak}
Left panel: Pressure dependence of the charge of a MF particle with $R=1~\mu m$
in the neon discharge of Ref.~\cite{KRZ05} (squares)~\cite{BFK08}. Middle and right panel:
Radius dependence of the charge of the MF particle for $p=50~Pa$ and $p=100~Pa$,
respectively. In all panels solid lines denote the (isothermal) charges deduced
from the surface model whereas dotted, dashed, and long-dashed lines are the charges
obtained from balancing on the grain surface $j_e^{\rm OML}$ with $j_i^{\rm OML}+j_i^{\rm CX}$
using, respectively, $\sigma_{\rm cx}=0.41\times 10^{-14}cm^2$~\cite{HES82},
which is the experimentally measured cross section at $0.12~eV$,
$\sigma_{\rm cx}=1.0\times 10^{-14}cm^2$, and $\sigma_{\rm cx}=2.0\times 10^{-14}cm^2$.
}
\end{center}
\end{figure}

That the charges obtained from the orbital motion limited flux balance are not 
too close to the experimental data is expected. But even the charge-exchange enhanced 
flux balance gives not particularly good results. This can be also seen 
in Fig.~\ref{Khrapak} where we analyze the charges of MF particles confined in the 
bulk of the neon discharge of Ref.~\cite{KRZ05}. 

The pressure dependence of the charge of a MF particle with $R=1~\mu m$ is shown 
in the left panel of Fig.~\ref{Khrapak}. Since the plasma parameters entering 
Eq.~(\ref{Zpfinal}) are known~\cite{KRZ05}, $T_p$ is again the only free parameter.
Fixing $T_p$ at a particular value gives the isothermal 
particle charges $Z_p(T_p)$ shown by the solid lines. From $Z_p(T_p)=Z_{exp}$ 
follows then the $T_p$ required to reproduce the experimental charge. The predicted 
increase of $T_p$ with pressure is again in accordance with the results of the 
calorimetric study of noble gas discharges presented in~\cite{SKD00}. 

In Fig.~\ref{Khrapak} we also plot the charges obtained from the charge-exchange 
enhanced flux balance condition. For $\sigma_{\rm cx}=2\times 10^{-14}~cm^2$
the agreement with the data is in fact quite good but this value of the cross 
section is almost five times larger than the experimentally measured value 
($\sigma_{\rm cx}=0.41\times 10^{-14}~cm^2$ at 
$0.12~eV$~\cite{HES82}). That the agreement with the experimental data is not too 
good can be also inferred from the radius dependence of $Z_p$ for $p=50~Pa$ and 
$p=100~Pa$ which we show, respectively, in the middle and the right panel of 
Fig.~\ref{Khrapak}. Clearly, the radius dependence of the grain charge seems
to be closer to the nonlinear dependence obtained from Eq.~(\ref{Zpfinal}) than
to the linear dependence resulting from the charge-exchange enhanced flux balance,
irrespective of the chosen value of the charge-exchange cross section.

\begin{table}[b]
\caption{\label{ArPara} Plasma parameters~\cite{TLA00}
used to obtain the results presented in Fig.~\ref{Tomme}.}
\begin{tabular}{llllll}
$p[Pa]$ & $n_e[10^9 cm^{-3}]$ & $kT_e[eV]$ & $kT_i[eV]$ & $V_{\rm rf}[V]$\\\hline
6.6 & 1.7 & 3.7 & 0.026 & 96.4\\
13.6 & 2.4 & 3.9 & 0.026 & 96.4
\end{tabular}
\end{table}

\begin{figure}[t]
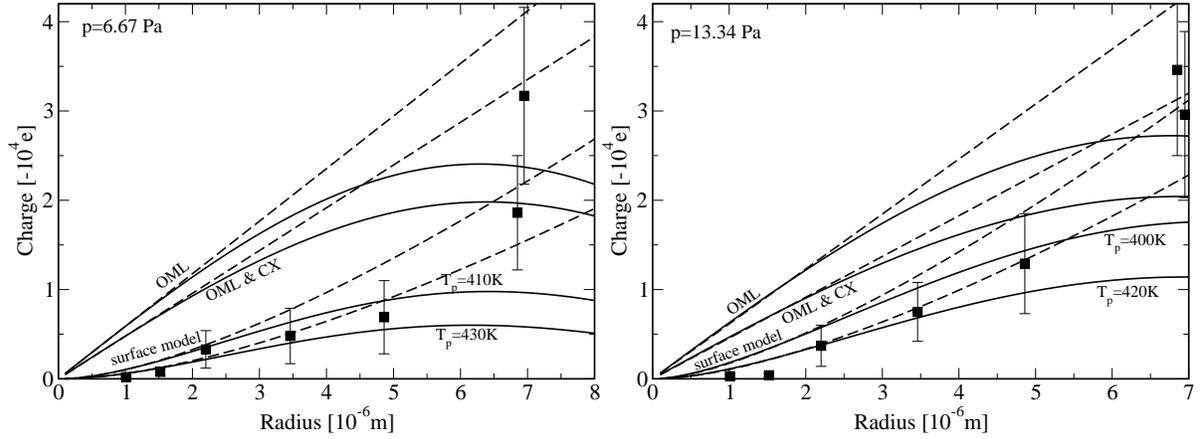

\begin{center}
\begin{minipage}{0.5\linewidth}
\includegraphics[width=\linewidth]{Fig6l.eps}
\end{minipage}\begin{minipage}{0.5\linewidth}
\includegraphics[width=\linewidth]{Fig6r.eps}
\end{minipage}
\caption{\label{Tomme}
Radius dependence of the charge of a MF particle in the sheath of
an argon discharge at $p=6.67~Pa$ (left panel)~\cite{BFK08} and $p=13.34~Pa$
(right panel). Squares are experimental data from~\cite{TAA00}
and solid and dashed lines give, respectively, the charges
obtained when the depletion of $n_e$ in the confining sheath
is included or not. For comparison we also show the charges
deduced from $j_e^{\rm OML}=j_i^{\rm OML}$ (OML) and from
$j_e^{\rm OML}=j_i^{\rm OML}+j_i^{\rm CX}$ (OML~$\&$~CX) with
$\sigma_{\rm cx}=0.72\times 10^{-14} cm^2$ which 
is the experimentally measured cross section at $0.1~eV$~\cite{HES82}.
}
\end{center}
\end{figure}

Finally, Fig.~\ref{Tomme}, showing for two different pressures the radius dependence 
of $Z_p$ for MF particles confined in the sheath of an argon discharge~\cite{TAA00}, 
the plasma parameters of which are given in table~\ref{ArPara}, provides additional 
support for our model. To approximately account for the
fact that particles with different radius experience different
plasma environments, we included the depletion of $n_e$ in
the sheath by replacing $n_e$ in $j_{e}^{\rm OML}$ by
$n_e\exp[e\Phi(z_{\rm eq}(R))/k_BT_e]$ with $\Phi(z)$ the sheath potential
and $z_{\rm eq}(R)$ the measured equilibrium position of the particle with
radius $R$~\cite{TAA00}. When the grains are not
too deep in the sheath ($R<5~\mu m$), we find for $p=6.67~Pa$ and $p=13.34~Pa$ 
excellent agreement with the data for, respectively, $T_p\simeq 410~K-430~K$ 
and $T_p=400~K-420~K$, although the particle temperatures are perhaps somewhat too 
high in view of the small amplitude of the rf voltage. Our approach fails, however, 
completely for $R>5~\mu m$. We attribute this to the ad-hoc modification of the electron
flux which may not capture the total electron flux close to the electrode. An improved 
treatment would calculate the electron flux self-consistently taking not only the depletion
of the electron density into account but also the flux due to sub-thermal secondary electrons 
from the electrode~\cite{SV03}. 

\section{Conclusions}

In this paper, we discussed the main assumptions underlying the surface model proposed 
in~\cite{BFK08} for the calculation of the charge of a dust particle immersed in a 
quiescent plasma and confronted the model with additional experimental data. 

The main hypothesis of the model, suggested by the analogy of charging of dust 
particles with physisorption of charged particles to the particles' surface, 
is that electrons and ions are, on a microscopic scale, spatially separated 
because the potential in which physisorption of electrons takes place is the 
attractive short-range polarization potential whereas for ions it is the 
attractive long-range Coulomb potential. For electrons the analogy may be obvious.
But for ions it may be not. However, without the grain no attractive long-range 
Coulomb potential would exist. Thus, the ion dynamics and kinetics in the 
vicinity of the grain is also a kind of surface physics although it occurs 
deep in the disturbed region of the grain and may be strongly affected by the 
plasma environment. 

Within the surface model, the grain charge and its partial screening can be calculated 
by balancing, on two different surfaces, the electron collection flux 
with the electron desorption flux and the ion collection flux with the ion desorption
flux. The charge of the grain is then given by the number of electrons quasi-bound 
in the short-range polarization potential of the grain whereas its partial screening 
is given by the number of ions quasi-bound in the long-range Coulomb potential. 

The grain temperature turns out to be an important parameter. Since, for the 
experiments we analyzed, the grain temperature has not been measured we used 
it as an adjustable parameter and obtained for physically meaningful particle 
temperatures excellent agreement with the experimentally measured grain charges. 
We challenge therefore experimentalists to simultaneously measure the grain charge
and the grain temperature. Our model could then be easily tested. 

The charges obtained from the surface model depend on the surface parameter 
$(s\tau)_e$ which is the product of the electron sticking coefficient with 
the electron desorption time. Both parameters depend on the inelastic, that 
is, the dynamic interaction between the electron and the grain. As discussed 
in~\cite{BFD09}, a rigorous calculation of $(s\tau)_e$ has to be 
based on a microscopic model for the electron-grain interaction taking 
elementary excitations of the grain into account. Quantum-kinetic equations 
have then to be solved to obtain $s_e$ and $\tau_e$ separately. For the product, 
however, a rough estimate, which turns out to be surprisingly good, can be 
obtained from the Lennard-Jones-Devonshire formula~(\ref{stau}) relating 
$(s\tau)_e$ to the particle temperature and the electron desorption energy. 
Measuring these two quantities directly would eliminate any free parameter
from our model. 

The surface model is a first attempt to treat plasma-controlled electron 
and ion fluxes and material-controlled plasma-wall interactions at the 
grain surface on an equal footing. Even in the present rudimentary form
the model performs better than approaches relying exclusively on an improvement 
of the plasma fluxes. Hence, if nothing else, it indicates that the charge a dust 
particle acquires in a plasma depends not only on the macroscopic plasma environment 
but also on microscopic processes on the surface occurring on a scale which is beyond 
the Boltzmann-Poisson description of the plasma-grain interaction. A quantitative theory 
of grain charging has to be therefore based on a systematic exploration of this ultimate 
boundary layer.

\begin{acknowledgement}
Support from the SFB-TR 24 ``Fundamentals of  Complex Plasmas'' is 
greatly acknowledged. F.~X.~B. is also grateful to M. Lampe for a 
particularly illuminating discussion. 
\end{acknowledgement}


\bibliographystyle{cpp}
\bibliography{DustParticle}

\end{document}